\documentclass{jpsj2}
\usepackage{color}

\def\runtitle{
BCS-BEC Crossover in Attractive Hubbard Model under Magnetic Field
}
\def\runauthor{
Atsushi Tsuruta, Satoshi Hyodo and Kazumasa Miyake
}

\title{
BCS-BEC Crossover in Two-Dimensional Attractive Hubbard Model 
under Magnetic Field
}
\author{
Atsushi Tsuruta$^{1}$, Satoshi Hyodo$^{1}$, and Kazumasa Miyake$^{2}$
}
\inst
{
$^{1}$Division of Materials Physics, Department of Materials Engineering Science \\
Graduate School of Engineering Science, Osaka University, Toyonaka, Osaka 560-8531, Japan\\
$^{2}$Toyota Physical and Chemical Research Institute, Nagakute, Aichi 480-1192, Japan
}

\recdate
{
\today
}

\abst
{
The Bardeen-Cooper-Schrieffer (BCS)-Bose-Einstein condensation (BEC) crossover in the 
two-dimensional attractive Hubbard model under the magnetic 
field is discussed at the half-filling at $T=0$ K on the basis of the formalism of 
Eagles and Leggett.  It is shown that the so-called
Fulde-Ferrel-Larkin-Ovchinikov (FFLO)-like state with a non-zero 
center-of-mass wave vector ${\bf q}\not=0$ is not stabilized in the weak-coupling (BCS) region, while such a state with ${\bf q}\not=0$ is stabilized against 
that with ${\bf q}=0$ even in a wide strong-coupling (BEC) region where 
di-fermion molecules are formed.  
The physical implication of this surprising result is discussed. 
}

\begin{document}
\sloppy
\maketitle
The problem of the Bardeen-Cooper-Schrieffer (BCS)-Bose-Einstein condensation (BEC) 
crossover has long been discussed after the BCS theory was established as the theory of 
superconductivity.  The description of the crossover between these ground states is 
relatively simple in both three~\cite{Wada,Eagle,Leggett} and 
two~\cite{Miyake,Randeria} dimensions. However, the crossover of the 
transition temperature $T_{\rm c}$ is much more strongly involved because 
one has to properly take into account the center-of-mass (COM) degrees of freedom 
of pairs, as first discussed by Nozi\`eres and Schmitt-Rink~\cite{N-SR}.  
In this decade, this problem has been revived in the context of research on 
cold atoms~\cite{Ohashi,Review,Tamaki}, after almost two decades since great interest 
was devoted to it in relation to high-$T_{\rm c}$ cuprate superconductors, which were thought to be in the crossover region owing to the shortness of 
the Cooper pair size~\cite{Micnas,Randeria2,SR-V-R,Tokumitu}.  

The purpose of this {\it Letter} is to clarify how the crossover of the ground 
state occurs under the magnetic field in the case of the attractive Hubbard 
model in two dimensions 
as a general problem that is not necessarily related to cold-atom systems.
Namely, we discuss the crossover of the so-called 
Fulde-Ferrel-Larkin-Ovchinikov (FFLO)-like state.  At first sight, the FFLO-like 
state is destabilized in the BEC limit where the tight di-fermion molecule is 
expected to be formed and to exhibit BEC with the zero COM wave vector at least 
in the free space.  However, it is a nontrivial problem whether the COM wave vector 
of the di-fermion molecule is zero in the attractive Hubbard model on the square lattice 
near the half-filling owing to the cooperative effects of the lattice periodicity and the 
magnetic field.  

The Hamiltonian of the attractive Hubbard model under the magnetic field used in 
the grand canonical ensemble is 
\begin{equation}
H=\sum_{\sigma}\left[-t\sum_{\langle i,j\rangle}
(c^{\dagger}_{i\sigma}c_{j\sigma}+{\rm h.c.})
-\sum_{i}(\sigma H+\mu)c^{\dagger}_{i\sigma}c_{i\sigma}\right]
-U\sum_{i}c^{\dagger}_{i\uparrow}c_{i\uparrow}c^{\dagger}_{i\downarrow}c_{i\downarrow},
\label{eq:1}
\end{equation}
where $\mu$ is the chemical potential and $\langle i,j\rangle$ means that 
the summation is taken over the nearest-neighbor pairs on the square lattice.  
In the $k$-representation, Eq. (\ref{eq:1}) is given in the form
\begin{equation}
H=\sum_{\sigma}\sum_{{\bf k}}(\xi_{\bf k}-\sigma H)
c^{\dagger}_{{\bf k}\sigma}c_{{\bf k}\sigma}
-\frac{U}{N_L}\sum_{{\bf q}}\sum_{{\bf k},{\bf k}'}
c^{\dagger}_{{\bf k}+{\bf q}/2\uparrow}c^{\dagger}_{-{\bf k}+{\bf q}/2\downarrow}
c_{-{\bf k}'+{\bf q}/2\downarrow}c_{{\bf k}'+{\bf q}/2\uparrow},
\label{eq:2}
\end{equation}
where $N_L$ is the number of lattice points,
and $\xi_{\bf k}$ is the kinetic energy of fermions measured from the 
chemical potential $\mu$, 
\begin{equation}
\xi_{\bf k}=-2t(\cos\,k_{x}a+\cos\,k_{y}a)-\mu,
\label{eq:3}
\end{equation}
where $a$ is the lattice constant.  
The mean-field Hamiltonian for Eq. (\ref{eq:2}) is given in a generalized BCS 
formula as discussed in the FFLO problem~\cite{FF,LO}:
\begin{eqnarray}
& &H_{\rm MF}=\sum_{\bf k}\biggl[(\xi_{{\bf k}+{\bf q}/2}-H)
c^{\dagger}_{{\bf k}+{\bf q}/2\uparrow}c_{{\bf k}+{\bf q}/2\uparrow}
+(\xi_{-{\bf k}+{\bf q}/2}+H)
c^{\dagger}_{-{\bf k}+{\bf q}/2\downarrow}c_{-{\bf k}+{\bf q}/2\downarrow}
\nonumber
\\
& &\qquad\qquad
-\Delta_{\bf q}^{*}
c_{-{\bf k}+{\bf q}/2\downarrow}c_{{\bf k}+{\bf q}/2\uparrow}
-\Delta_{\bf q}
c^{\dagger}_{{\bf k}+{\bf q}/2\uparrow}c^{\dagger}_{-{\bf k}+{\bf q}/2\downarrow}
+\Delta_{\bf q}
\langle 
c^{\dagger}_{{\bf k}+{\bf q}/2\uparrow}c^{\dagger}_{-{\bf k}+{\bf q}/2\downarrow}
\rangle
\biggr],
\label{eq:4}
\end{eqnarray}
where $\langle\cdots\rangle$ is the grand canonical ensemble average 
concerning the mean-field Hamiltonian Eq. (\ref{eq:4}) itself, and 
the gap $\Delta_{\bf q}$ satisfies the self-consistent equation 
\begin{equation}
\Delta_{\bf q}=\frac{U}{N_L}\sum_{\bf k}\langle 
c_{-{\bf k}+{\bf q}/2\downarrow}c_{{\bf k}+{\bf q}/2\uparrow}\rangle.
\label{eq:5}
\end{equation}

\begin{figure}[h]
\begin{center}
\rotatebox{0}{\includegraphics[width=0.75\linewidth]{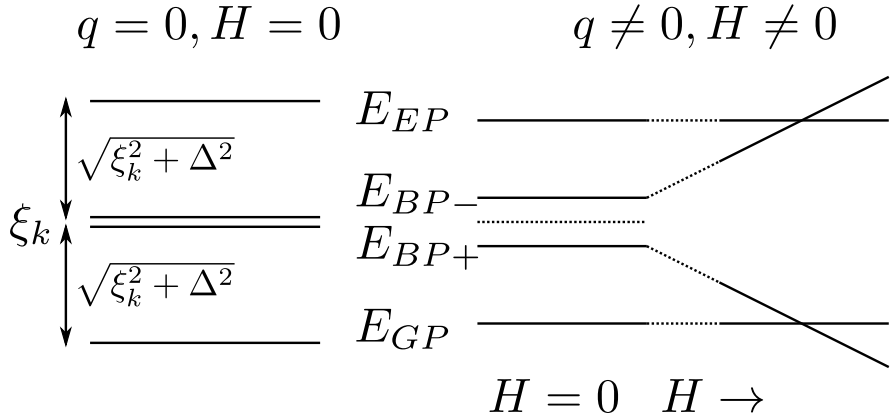}}
\caption{Energy level scheme, Eqs. (\ref{eq:6})$-$(\ref{eq:8b}), under 
the magnetic field $H$ in the Hilbert space with a fixed {\bf k} and a fixed {\bf q}. 
}
\label{Fig:1}
\end{center}
\end{figure}

The eigenvalue problem of this Hamiltonian is solved independently for each 
wave vector {\bf k} and the COM wave vector {\bf q} following the 
method adopted by Leggett in Ref. \citen{LeggettRMP}.  In the Hilbert space with 
a fixed {\bf k} and a fixed {\bf q}, eigenvalues are given as 
\begin{eqnarray}
& &E_{\rm GP}={1\over 2}\left[
(\xi_{{\bf k}+{\bf q}/2}+\xi_{-{\bf k}+{\bf q}/2})
-\sqrt{(\xi_{{\bf k}+{\bf q}/2}+
\xi_{-{\bf k}+{\bf q}/2})^{2}+4\Delta_{\bf q}^{2}}\right]\equiv 
\Xi_{{\bf k},{\bf q}},
\label{eq:6}
\\
& &E_{\rm EP}={1\over 2}\left[
(\xi_{{\bf k}+{\bf q}/2}+\xi_{-{\bf k}+{\bf q}/2})
+\sqrt{(\xi_{{\bf k}+{\bf q}/2}+
\xi_{-{\bf k}+{\bf q}/2})^{2}+4\Delta_{\bf q}^{2}}\right],
\label{eq:7}
\\
& &E_{\rm BP}^{+}=\xi_{{\bf k}+{\bf q}/2}-H,
\label{eq:8a}
\\
& &E_{\rm BP}^{-}=\xi_{-{\bf k}+{\bf q}/2}+H.
\label{eq:8b}
\end{eqnarray}
The energy level scheme under the magnetic field $H$ is shown in Fig.\ \ref{Fig:1}.  
The ground state in this restricted Hilbert space is the ground-pair (GP) state 
if $(\xi_{-{\bf k}+{\bf q}/2}-H)>\Xi_{{\bf k},{\bf q}}$, and the broken-pair 
(BP$^{+}$) state with the up spin if $(\xi_{-{\bf k}+{\bf q}/2}-H)<\Xi_{{\bf k},{\bf q}}$. 
Taking this fact into account, the gap equation at $T=0$ is given as 
\begin{equation}
\Delta_{\bf q}=\frac{U}{N_L}\sum_{\bf k}
{\Delta_{\bf q}\over \sqrt{(\xi_{{\bf k}+{\bf q}/2}+
\xi_{-{\bf k}+{\bf q}/2})^{2}+4\Delta_{\bf q}^{2}}}
\theta(\xi_{{\bf k}+{\bf q}/2}-H-\Xi_{{\bf k},{\bf q}}),
\label{eq:9}
\end{equation}
where $\theta$ is the Heaviside function.  
Similarly, the total number 
of fermions per site at $T=0$ is given as 
\begin{eqnarray}
& &N=\frac{1}{N_L}\sum_{\bf k}\left[1-{\xi_{{\bf k}+{\bf q}/2}+\xi_{-{\bf k}+{\bf q}/2}
\over \sqrt{(\xi_{{\bf k}+{\bf q}/2}+
\xi_{-{\bf k}+{\bf q}/2})^{2}+4\Delta_{\bf q}^{2}}}
\right]
\theta(\xi_{{\bf k}+{\bf q}/2}-H-\Xi_{{\bf k},{\bf q}})
\nonumber
\\
& &\qquad
+\frac{1}{N_L}\sum_{\bf k}
\left[1-\theta(\xi_{{\bf k}+{\bf q}/2}-H-\Xi_{{\bf k},{\bf q}})\right]
\theta(H-\xi_{{\bf k}+{\bf q}/2}).
\label{eq:10}
\end{eqnarray}
Equation (\ref{eq:10}) determines the chemical potential $\mu$ as a function of U.  Then, 
Eqs. (\ref{eq:9}) and (\ref{eq:10}) should be solved simultaneously as 
in the theoretical framework of Eagles\cite{Eagle} and Leggett~\cite{Leggett}.  
With the use of thus-determined $\Delta_{\bf q}$ and $\mu$, the ground-state energy per site with a fixed {\bf q} is given as   
\begin{eqnarray}
& &E_{0}=\frac{1}{N_L}
\sum_{\bf k}\left[\Xi_{{\bf k},{\bf q}}
+{\Delta_{\bf q}^{2}\over \sqrt{(\xi_{{\bf k}+{\bf q}/2}+
\xi_{-{\bf k}+{\bf q}/2})^{2}+4\Delta_{\bf q}^{2}}}
\right]\theta(\xi_{{\bf k}+{\bf q}/2}-H-\Xi_{{\bf k},{\bf q}})
\nonumber
\\
& &\qquad
+\frac{1}{N_L}
\sum_{\bf k}(\xi_{{\bf k}+{\bf q}/2}-H)
\left[1-\theta(\xi_{{\bf k}+{\bf q}/2}-H-\Xi_{{\bf k},{\bf q}})\right]
\theta(H-\xi_{{\bf k}+{\bf q}/2}).
\label{eq:11}
\end{eqnarray}
The true ground-state energy is determined so that the energy given by Eq. (\ref{eq:11}) 
becomes minimum with the relaxation of the COM wave vector ${\bf q}$.  

\begin{figure}[h]
\begin{center}
\rotatebox{0}{\includegraphics[width=1.0\linewidth]{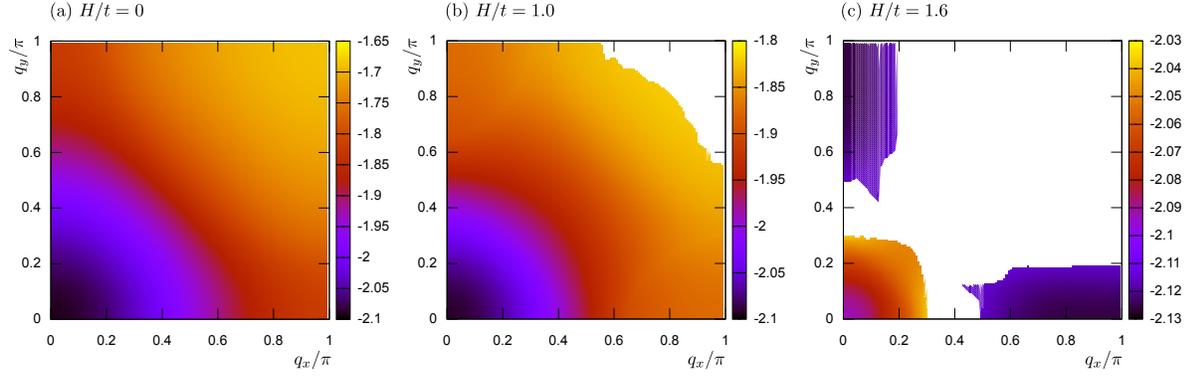}}
\caption{(Color) Distribution of the ground-state energy, Eq. (\ref{eq:11}), 
in the space of COM wave vector {\bf q} for the attractive interaction 
$U=6t$. The values of magnetic field are (a) $H/t=0$, (b) $H/t=1.0$, and 
(c) $H/t=1.6$. 
}
\label{Fig:2}
\end{center}
\end{figure}

We solve the self-consistent equations (\ref{eq:9}) and (\ref{eq:10}) numerically 
by dividing the first Brillouin zone 
up to
1000$\times$1000 meshes; 
we then determine the ground-state energy given by Eq. (\ref{eq:11}) for a fixed COM wave vector {\bf q}.  
Then, we seek the wave vector {\bf q} that minimizes $E_{0}$ given by Eq. (\ref{eq:11}). The filling of electrons is restricted within the case of half-filling, 
otherwise stated explicitly.
We have verified that the results are essentially independent of the mesh size of the first Brillouin zone even if the mesh size decreases from 900$\times$900 to 500$\times$500.
Figure\ \ref{Fig:2} shows some examples of the distribution 
of the ground-state energy $E_{0}$ in the space of the COM wave vector {\bf q} for $U=6t$ 
and $H/t=0$ [Fig.\ \ref{Fig:2}(a)], $H/t=1.0$ [Fig.\ \ref{Fig:2}(b)], and 
$H/t=1.6$ [Fig.\ \ref{Fig:2}(c)].

\begin{figure}[h]
\begin{center}
\rotatebox{0}{\includegraphics[width=0.7\linewidth]{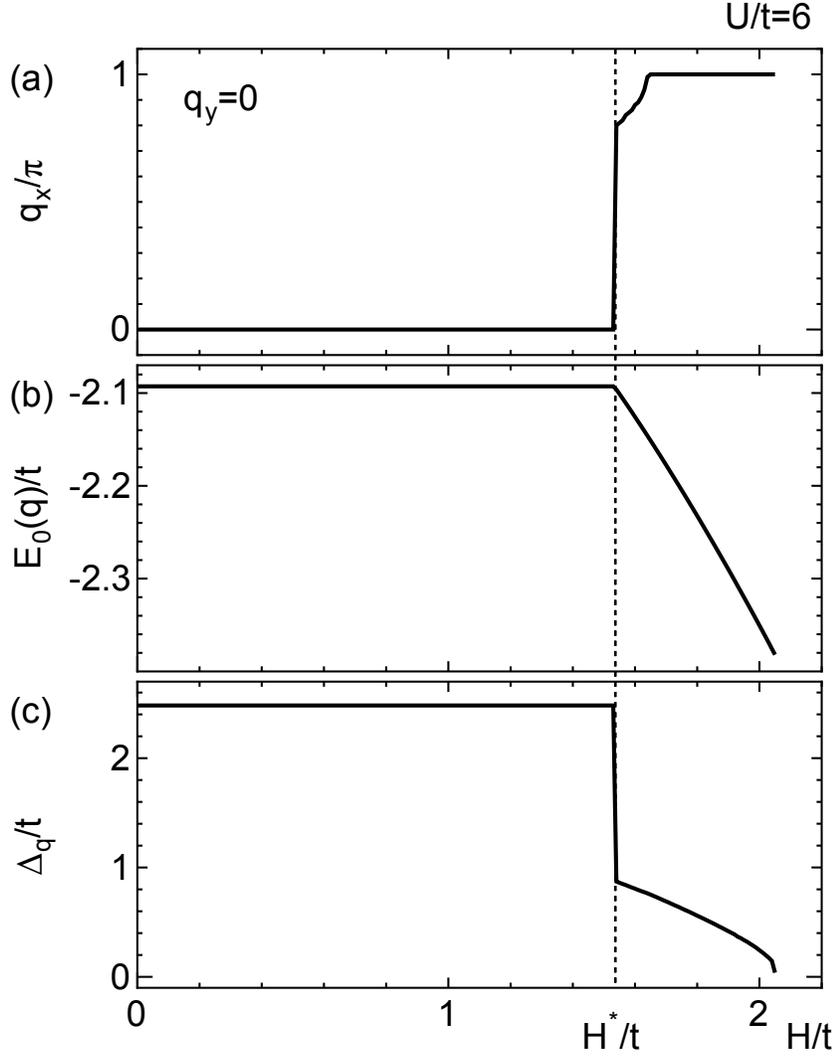}}
\caption{
(a) $q_{x}$, $x$-component of COM wave vector, 
(b) ground-state energy $E_{0}/t$ given by Eq. (\ref{eq:11}), and 
(c) superconducting gap $\Delta_{q}/t$ as a function of attractive 
interaction $H/t$.  
}
\label{Fig:3}
\end{center}
\end{figure}

In the case of $U/t=6$, for the magnetic field $0<H/t<2.05$, the COM wave vector minimizing 
the ground-state energy, Eq. (\ref{eq:11}), is {\bf q}=($q_{x},q_{y}=0$) or 
{\bf q}=($q_{x}=0,q_{y}$).  Such $q_{x}$ is drawn in Fig.\ \ref{Fig:3}(a), together 
with the ground-state energy $E_{0}/t$ [Fig.\ \ref{Fig:3}(b)], and 
the superconducting gap $\Delta_{q}/t$ [Fig.\ \ref{Fig:3}(c)].  
It is remarked that $q_{x}=0$ holds up to $H/t=H^{*}/t\simeq 1.53$ where a first order transition 
occurs from the state with $q_{x}=0$ to that with $q_{x}\not=0$.  The transition 
to the normal state at $H/t\simeq 2.05$ is a second order.  

\begin{figure}[h]
\begin{center}
\rotatebox{0}{\includegraphics[width=0.8\linewidth]{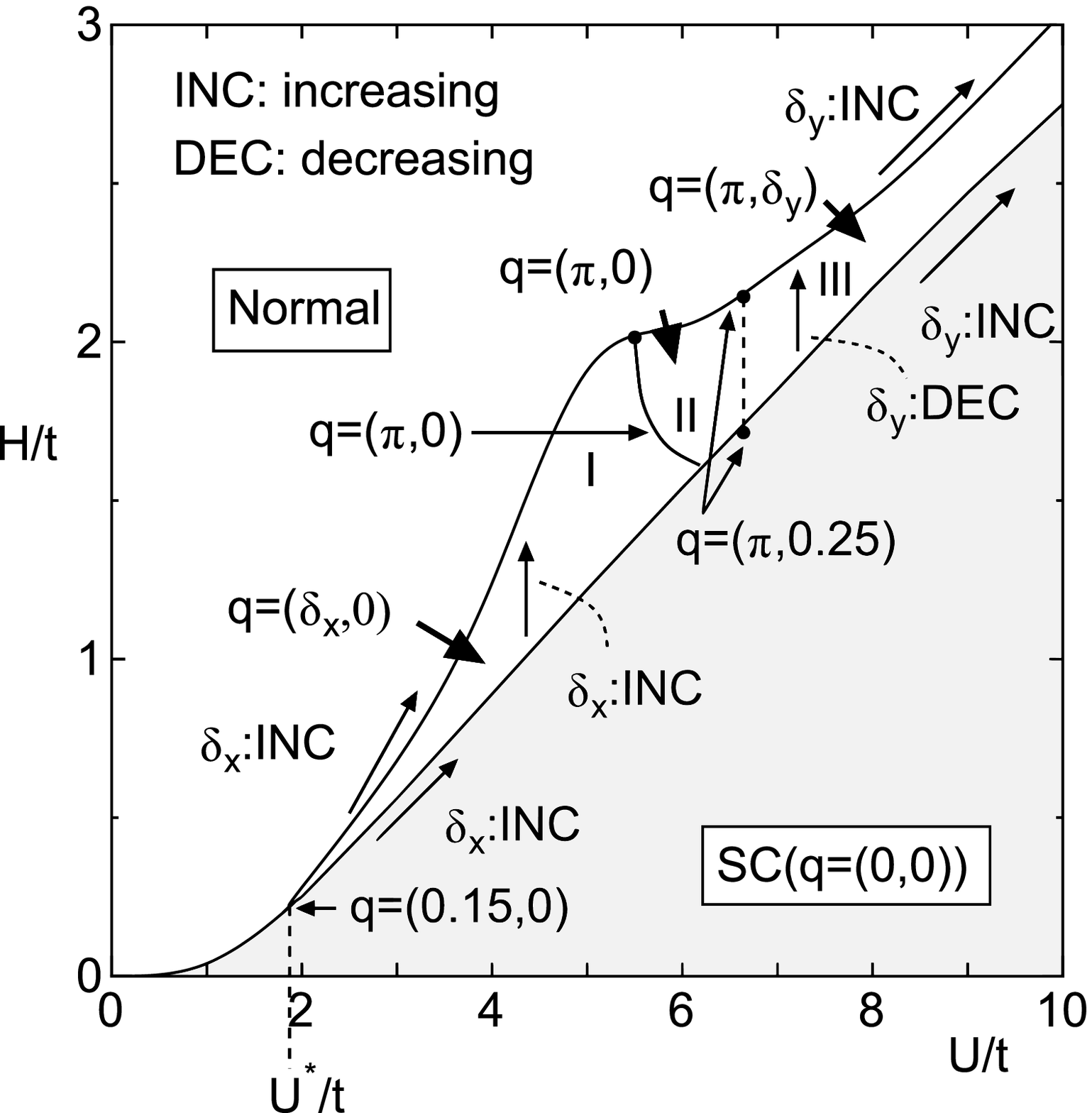}}
\caption{
Phase diagram of the ground state in $U/t$-$H/t$ plane. The solid and dashed 
lines represent the second-order and first-order transitions, 
respectively. 
The wave vector $q$ in regions I, II, and III are
$(\delta_x,0)$, $(\pi,0)$ and $(\pi,\delta_y)$, respectively.
}
\label{Fig:4}
\end{center}
\end{figure}

We have searched for the COM wave vector minimizing the ground-state energy, 
Eq. (\ref{eq:11}), for a series of sets of $U/t$ and $H/t$. 
The resultant phase diagram in the $U/t$-$H/t$ plane is shown in Fig.\ \ref{Fig:4}
where the second-order transition is indicated by the solid lines and the 
first-order transition by the dashed lines.  
It is remarkable that the FFLO-like state with ${\bf q}\not=(0,0)$ does not exist in the 
region of attractive interaction $U<U^{*}$, $U^{*}/t\simeq2.0$, which is a new aspect 
of issues on the possibility of the FFLO-like state in lattice systems.  
Indeed, in the ground state of the continuum model, there always exists the FFLO-like state between 
the conventional pairing state with {\bf q}=(0,0) and the normal state~\cite{FF,LO,FFLOREV}.  

\begin{figure}[h]
\begin{center}
\rotatebox{0}{\includegraphics[width=0.9\linewidth]{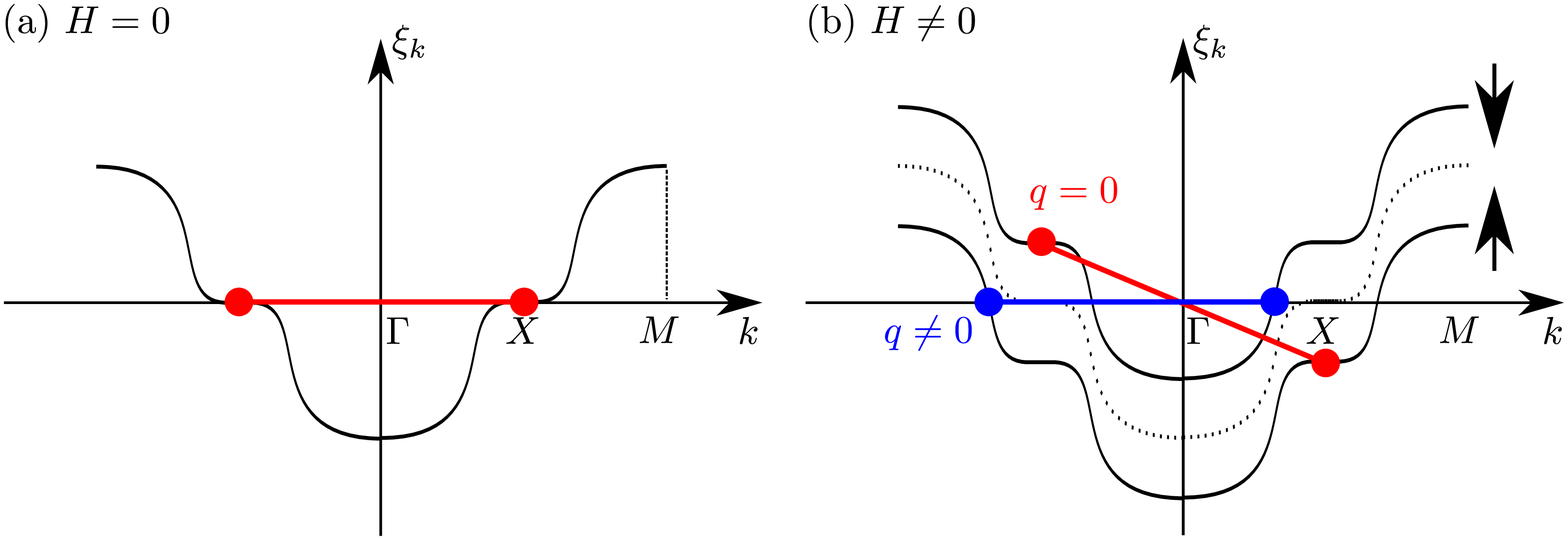}}
\caption{
(Color) (a) Schematic band dispersion along the wave vector $\Gamma$-$X$-$M$ 
[$(k_{x},k_{y})$: $(0,0)$-$(\pi,0)$-$(\pi,\pi)$] without 
magnetic field, i.e., $H=0$.  
The red line indicates the path of Cooper pair formation with ${\bf q}=0$.   
(b)
Schematic band dispersion along the wave vector $\Gamma$-$X$-$M$ 
[$(k_{x},k_{y})$: $(0,0)$-$(\pi,0)$-$(\pi,\pi)$] with 
magnetic field, i.e., $H\not=0$.  
The red line indicates the path of Cooper pair formation with ${\bf q}=0$, 
while the blue line indicates that with ${\bf q}\not=0$.  
}
\label{Fig:5}
\end{center}
\end{figure}

The reason why the conventional Cooper pair with ${\bf q}=0$ is stable against 
the magnetic field up to $H=H^{*}$ is understood as follows: In the case of $H=0$, 
Cooper pair formation is promoted by the diverging density of states due to 
the van Hove singularity in addition to the usual Cooper effect, as shown in 
Fig.\ \ref{Fig:5}(a).  Even under the 
weak magnetic field, this additional effect due to the van Hove singularity works 
to compensate for the effect of magnetic fields destabilizing the pair with {\bf q}=0 
as shown in Fig.\ \ref{Fig:5}(b).  We have verified that the FFLO-like state 
recovers for which the van Hove singularity shifts sufficiently from the Fermi surface.  

\begin{figure}[h]
\begin{center}
\rotatebox{0}{\includegraphics[width=0.8\linewidth]{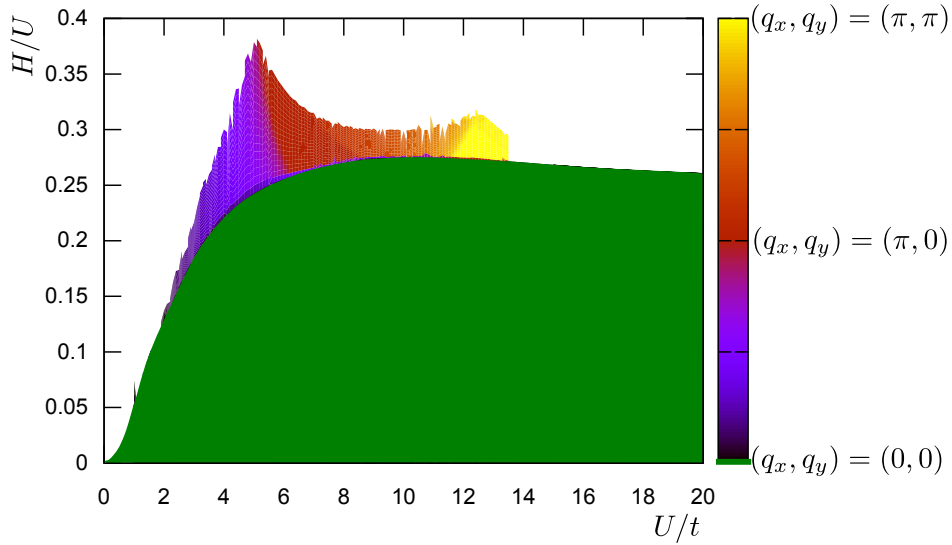}}
\caption{(Color) Phase diagram of ground state in $H/U$-$U/t$ plane. 
}
\label{Fig:6}
\end{center}
\end{figure}

\begin{figure}[h]
\begin{center}
\rotatebox{0}{\includegraphics[width=0.6\linewidth]{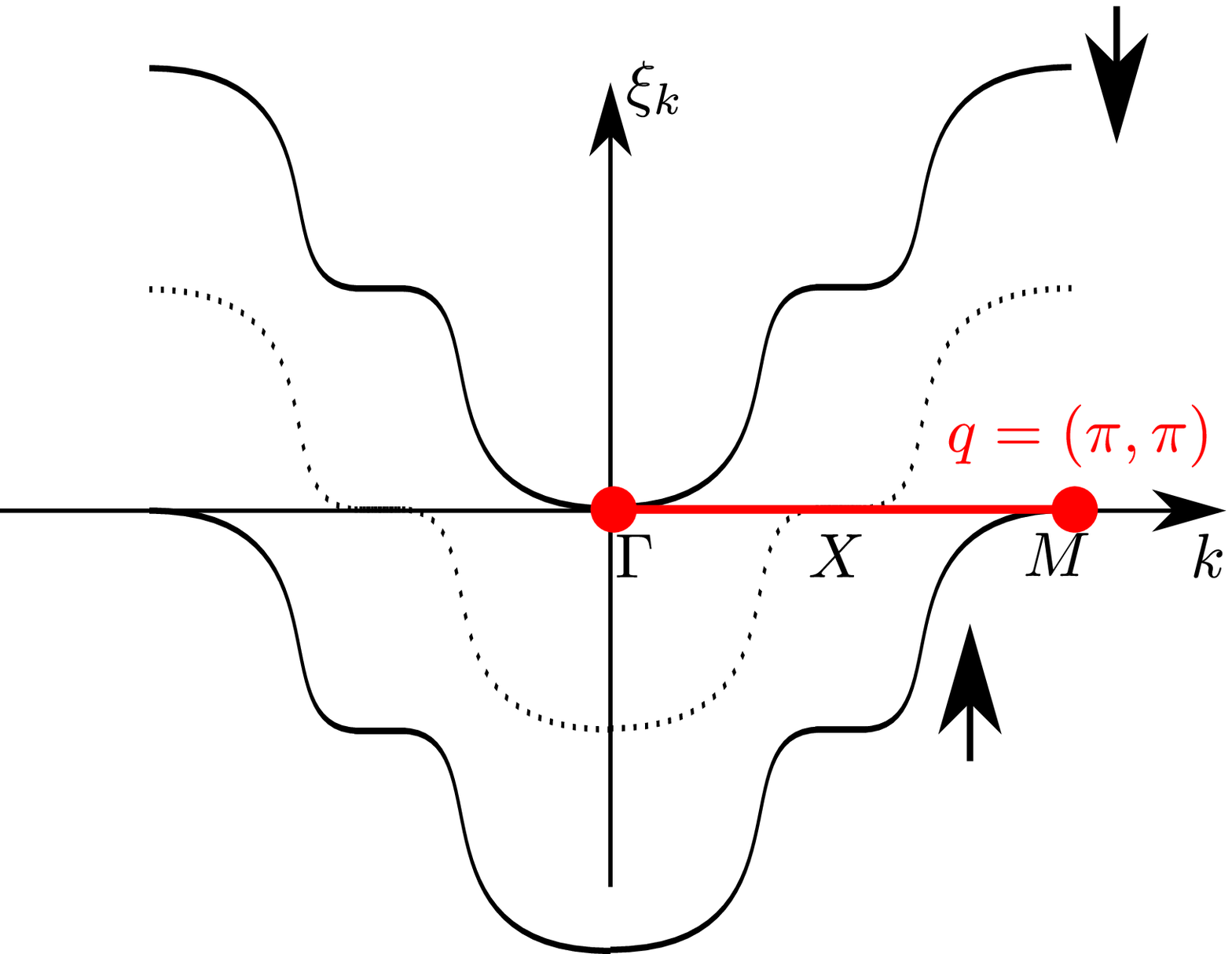}}
\caption{(Color) Dispersion of electrons along the path connecting $\Gamma$-X-M points. 
}
\label{Fig:7}
\end{center}
\end{figure}

The phase diagram in a wider region of parameters $H/U$-$U/t$ is shown 
in Fig.\ \ref{Fig:6}.  Note that there exists a region where 
the pairing with {\bf q}=($\pi,\pi$) is stabilized in the strong-coupling 
region $U/t\sim 12$ and $H/U\sim 0.28$.  The attractive interaction is larger than 
the half bandwidth $W/t=4$ so that the BEC region is realized.  
At first sight, this is somewhat surprising because it seems difficult for a tightly 
bound ``di-electronic molecule'' to acquire the COM wave vector.
However, 
it turns out to be rather natural considering the dispersion of electrons 
under the strong magnetic field larger than W, the half bandwidth, as shown in Fig.\ \ref{Fig:7}.  
Namely, the ``molecule'' is expected to form between the electron around the $\Gamma$ point 
{\bf k}=(0,0) with the down spin (the bottom of the minority band) and that around the M point 
{\bf k}=($\pi,\pi$) with the up spin (the top of the majority band), 
because the energy difference between the up-spin band and the down-spin band 
takes a minimum for such a combination of {\bf k} points, giving {\bf q}=($\pi,\pi$).  
The FFLO-like state does not exist in the weak-coupling region even in the half-filling case and the case in which the next-nearest-neighbor hopping is finite.
Details of effects of the next-nearest-neighbor hopping will be discussed elsewhere.

The Shiba transformation is defined as\cite{Shiba}
\begin{eqnarray}
&&b_{i\uparrow}^+ = c_{i\uparrow}^+\nonumber\\
&&b_{i\downarrow}^+={\rm e}^{ {\rm i}{\bf Q}_0\cdot {\bf R}_i}c_{i\downarrow}\nonumber\\
&&{\bf Q}_0=(\pi, \pi),
\end{eqnarray}
where ${\bf R}_i$ is the position vector in unit of the lattice constant $a$ of the simple square lattice.
By this transformation, the attractive Hubbard model Eq. (1) is transformed into the repulsive Hubbard model.
Adding the magnetic field in Eq. (1) corresponds to changing the chemical potential  (the particle number density) in the repulsive Hubbard model. 
In the transformed world, the BCS state and charge density wave (CDW) state in Eq. (1) are transformed to the transverse spin density wave (SDW) state and longitudinal SDW state, respectively, in the transformed repulsive Hubbard model. 
The SU(2) symmetry in the repulsive Hubbard model is saved even if we change the chemical potential.
Therefore, the degeneracy of the BCS state and CDW state is saved under the magnetic field in the attractive Hubbard model.
Here, we note that the charge density wave (CDW) state is degenerate with the BCS state in the system described by the model Hamiltonian Eq. (1).
Therefore, one might wonder how the CDW state is influenced by the magnetic field.
This issue is clarified by analysis using the so-called Shiba transformation.
In this sense, the magnetic field affects the CDW state in the same manner as in the BCS state.
Namely, the incommensurate component in CDW is expected to be induced by the magnetic field.
The FFLO-like state with ${\bf q}=(\pi,\pi)$, which is the yellow region in Fig.\ \ref{Fig:6} corresponds to the Nagaoka ferromagnetic ordered state in the repulsive Hubbard model.

Hirsch\cite{Hirsch} and many theorists investigated the SDW state in the repulsive Hubbard model.
Although they did not investigate detailed properties of the SDW state,
there exist differences between the results in Ref.\citen{Igoshev} for the repulsive Hubbard model in the mean field approximation and our results for the attractive Hubbard model with the same mean field approximation.
For example, 
in the repulsive Hubbard model, the incommensurate SDW state exists even in the weak-coupling region, and the SDW state with ${\bf q}=(Q,Q)\,(0<Q<\pi)$ can be stable near the half filling.
On the other hand, in the attractive Hubbard model, the FFLO-like state does not exist in the weak-coupling region, and the FFLO-like state with ${\bf q}=(\pi-Q,\pi-Q)\,(0<Q<\pi)$, which corresponds to ${\bf q}=(Q,Q)\,(0<Q<\pi)$ in the repulsive Hubbard model through the Shiba transformation, is not the ground state in the entire region of the $U-H$ phase diagram.
The reason for this difference is not clear at the moment.
It is suggestive to note that the theory of Nozi\`ere and Schmitt-Rink for the transition temperature $T_{\rm c}$~\cite{N-SR} when applied to the attractive Hubbard model is not equivalent to $T_{SDW}$, the SDW transition temperature,
but the equivalency is recovered in the fluctuation exchange (FLEX) approximation, which is applied to the thermodynamic potential treated by Nozi\`ere and Schmitt-Rink.\cite{Tamaki2}
This implies that the equivalency between the attractive and repulsive Hubbard models is not always maintained if some sort of approximation is introduced.

In conclusion, we have investigated the BCS-BEC crossover in the attractive Hubbard model 
on the square lattice 
under the magnetic field at the half-filling at $T=0$ K on the basis of the formalism of 
Eagles and Leggett.  It has been shown that the so-called FFLO-like state with a nonzero 
center-of-mass wave vector ${\bf q}\not=0$ is not stabilized in the weak-coupling (BCS) region, while such a state with ${\bf q}\not=(0,0)$ is stabilized against 
that with ${\bf q}=0$ even in a wide strong-coupling (BEC) region.  
In particular, {\bf q}=($\pi,\pi$) in the strong coupling limit $U\gg W$. 

\section*{Acknowledgments}
One of the authors (K.M.) has benefited from conversations on BCS-BEC crossover 
with H. Tamaki in the very early stage of this work.  
This work was supported by
a Grant-in-Aid for Scientific Research on Innovative Areas
``Topological Quantum Phenomena'' (No. 22103003)
from the Ministry of Education, Culture, Sports, Science 
and Technology and by a Grant-in-Aid for Scientific Research (No. 25400369) 
from the Japan Society for the Promotion of Science.

\end{document}